\documentclass[conference]{IEEEtran}
\IEEEoverridecommandlockouts
\usepackage{cite}
\usepackage{amsmath,amssymb,amsfonts}
\usepackage{algorithmic}
\usepackage{graphicx}
\usepackage{textcomp}
\usepackage{xcolor}
\usepackage{subcaption}
\usepackage{hyperref}
\usepackage{booktabs}
\usepackage{array}
\usepackage{tcolorbox}
\usepackage{enumitem,kantlipsum}
\definecolor{darkpastelgreen}{rgb}{0.01, 0.75, 0.24}

\def\BibTeX{{\rm B\kern-.05em{\sc i\kern-.025em b}\kern-.08em
    T\kern-.1667em\lower.7ex\hbox{E}\kern-.125emX}}

\begin{document}

\title{A Taxonomy of Error Sources in HPC I/O Machine Learning Models}



\author{
\IEEEauthorblockN{
    Mihailo Isakov\IEEEauthorrefmark{1},
    Mikaela Currier\IEEEauthorrefmark{1},
    Eliakin del Rosario\IEEEauthorrefmark{1},
    Sandeep Madireddy\IEEEauthorrefmark{2},\\
    Prasanna Balaprakash\IEEEauthorrefmark{2},
    Philip Carns\IEEEauthorrefmark{2},
    Robert B. Ross\IEEEauthorrefmark{2},
    Glenn K. Lockwood\IEEEauthorrefmark{3},
    Michel A. Kinsy\IEEEauthorrefmark{1}
}
\IEEEauthorblockA{
    \IEEEauthorrefmark{1}~\textit{Secure, Trusted, and Assured Microelectronics (STAM) Center}\\
    \textit{Ira A. Fulton Schools of Engineering, Arizona State University},  
    Tempe, AZ 85281\\
    \{misakov1,mkinsy\}@asu.edu
}
\IEEEauthorblockA{
    \IEEEauthorrefmark{2}~\textit{Argonne National Laboratory}, Lemont, IL 60439\\
    smadireddy@anl.gov, \{pbalapra,carns,rross\}@mcs.anl.gov
}
\IEEEauthorblockA{
    \IEEEauthorrefmark{3}~\textit{Lawrence Berkeley National Laboratory}, Berkeley, CA 94720\\
    glock@lbl.gov
}
}

\maketitle

\begin{abstract}
    I/O efficiency is crucial to productivity in scientific computing, but the increasing complexity of the system and the applications makes it
difficult for practitioners to understand and optimize I/O behavior at scale. Data-driven machine learning-based I/O
throughput models offer a solution: they can be used to identify bottlenecks, automate I/O tuning, or optimize job
scheduling with minimal human intervention. Unfortunately, current state-of-the-art I/O models are not robust enough for
production use and underperform after being deployed.

We analyze multiple years of application, scheduler, and storage system logs on two leadership-class HPC platforms to
understand why I/O models underperform in practice. We propose a taxonomy consisting of five categories of I/O modeling
errors: poor application and system modeling, inadequate dataset coverage, I/O contention, and I/O noise. We
develop litmus tests to quantify each category, allowing researchers to narrow down failure modes, enhance I/O
throughput models, and improve future generations of HPC logging and analysis tools.

\end{abstract}


\vspace{-0.5cm}
\section{Introduction}
As scientific applications push to leverage ever more capable computational platforms, there is a critical need to
identify and address bottlenecks of all types. 
Due to the large data movements in these
applications, the I/O subsystem is often a major source of performance bottlenecks, and it is common for applications to
attain only a small fraction of the peak I/O rates~\cite{luu:behavior}.  These performance problems can severely limit the
scalability of applications and are difficult to detect, diagnose, and fix.  Data-driven machine learning-based models
of I/O throughput can help practitioners understand application bottlenecks (e.g.,~\cite{isakov_sc20, moana,
10.1007/978-3-319-92040-5_10, 10.1145/3337821.3337922, isakov_ross20, 10.1145/3369583.3392678}), and have the potential to
automate I/O tuning and other tasks.  However, current machine learning-based I/O models are not robust enough for production
use~\cite{isakov_ross20}.
A thorough investigation of \emph{why} these models underperform when deployed on high
performance computing (HPC) systems will provide key insights and guidance on how to address their shortcomings. The
goal of our study is to help machine learning (ML)-driven I/O modeling techniques make the transition from theory to
practice.  

There are several reasons why machine learning-based I/O models underperform when deployed: poor modeling choices~\cite{isakov_sc20,
10.1145/3369583.3392678}, concept drift in the data~\cite{10.1145/3337821.3337922}, and weak generalization
~\cite{isakov_ross20},
among others. I/O models are often opaque, and there is no established methodology for diagnosing the root
cause of model errors. In this work, we present a taxonomy of ML-based I/O modeling errors, as shown in
Figure~\ref{fig:teaser}. Through this taxonomy, we show that I/O throughput prediction errors can be separated and
quantified into five error classes: inadequate (1) application and (2) system models, (3) novel application or system
behaviors, (4) I/O contention and (5) inherent noise. 
For each class, we present data-driven litmus tests that estimate the portion of
modeling error caused by that class. The taxonomy enables independent study of each source of error 
and prescribes appropriate ML techniques to tackle the underlying sources of error. 

Our contributions in this work are as follows:
\begin{enumerate}[leftmargin=*]
    \item We introduce a taxonomy of ML-based I/O throughput modeling errors which consists of five causes.

    \item We show that choice of machine learning model type, scaling model size, and hyperparameter tuning cannot reduce all potential errors. 
        We present two litmus tests that quantify error due to poor application and
        system modeling.

    \item We present a litmus test that estimates what portion of error is caused by
    rare jobs with previously unseen behavior, and apply uncertainty quantification methods to classify those jobs as
        out-of-distribution jobs.

    \item We present a method for quantifying the impact of I/O contention and noise on I/O throughput, which (1)
        defines a fundamental limit in how accurate ML models can become,
        and (2) gives HPC system users and
        administrators a practical estimate of the I/O throughput variance they should expect. We show that underlying
        system noise is the dominant source of errors, and not poor modeling or lack of application or system data. 

    \item We present a framework for how the taxonomy is practically applied to new systems and evaluate it on two
        leadership-class supercomputers: Argonne Leadership Computing Facility (ALCF) Theta and National Energy
        Research Scientific Computing Center (NERSC) Cori.
\end{enumerate}


\vspace{-0.3cm}
\section{Related work}\label{sec:related}
In recent years, automating HPC I/O system analysis through ML has received significant attention, with two prominent
directions: 
(1) workload clustering to better understand groups of HPC jobs and automate handling of whole groups, and 
(2) I/O subsystem modeling and make predictions of HPC job I/O time, I/O throughput, optimal scheduling, etc. 
Clustering HPC job logs has been explored in~\cite{isakov_sc20, gauge, taxonomist} with the
goal of better understanding workload distribution, scaling I/O expert effort more efficiently, and revealing hidden
trends and I/O workload patterns. ML-based modeling has been used for predicting I/O
time~\cite{10.1007/978-3-319-92040-5_10}, I/O throughput~\cite{10.1145/3369583.3392678, isakov_sc20}, optimal filesystem
configuration~\cite{8752835, capes}, as well as for building black boxes of I/O subsystems in order to apply ML model
interpretation techniques~\cite{isakov_sc20}. While there have been some attempts at creating analytical models of I/O
subsystems~\cite{10.1109/CLUSTER.2015.29}, most attempts are data-driven, and rely on HPC system logs to create models
of I/O \cite{luu:behavior, 10.1145/3369583.3392678, 10.1007/978-3-319-92040-5_10, isakov_sc20, Wang2018IOMinerLA}. 
Although the challenges of developing accurate machine learning models are well known, the nature of the domain requires
special consideration: I/O subsystems have to service multiple competing jobs, their configuration evolves over time,
they have periods of increased variability, they experience occasional hardware faults, etc.~\cite{1526010,
10.1109/SC.2018.00077, costa1}. Diagnosing this I/O variability, where the performance of a job depends on external factors to
the job itself has been extensively studied~\cite{moana, 1526010, 10.1145/3322789.3328743,
10.1007/978-3-319-92040-5_10, costa1}. Finally, the deployment of I/O models has been shown to require special consideration as
these models often significantly underperform on new applications~\cite{10.1145/3337821.3337922, isakov_ross20}. 
While different sources of model error have been studied individually, no prior work characterizes the relative impact
of different sources of error on model accuracy.

\section{Modeling HPC applications and systems}\label{sec:system}
The behavior of an HPC system is governed by both complex rules and inherent noise. By formalizing the system as a
mathematical function (or, more generally, a stochastic process) with its inputs and outputs, the process may be decomposed into
smaller components more amenable to analysis. The I/O throughput of a system running specific sets of applications may
be treated as a data-generating process from which I/O throughput measurements are drawn. While building a perfect model
of an HPC system may not be possible, it is useful to understand the inputs to the `true' process and the process's
functional properties. The theoretical model of the process must include all causes that might affect a real HPC system,
such as: how well a job uses the system, hardware and software configurations over the life of the system, resource
contention between concurrent jobs, inherent application-specific and system noise, as well as application-specific
noise sensitivity. Although many of these aspects are not observable in practice, the true
data-generating process takes these aspects into account.

We adapt the global modeling formulation from Madireddy et al.~\cite{10.1007/978-3-319-92040-5_10}, and formulate I/O
throughput $\phi(j)$ of a job $j$ as:
\begin{equation} \label{eq:global_model}
    \phi(j) = f(j, \zeta, \omega)
\end{equation}
Here, $\zeta$ represents system state (e.g., filesystem health, system configuration, node availability, etc.) and
system behavior (e.g., the behavior of other applications co-located with the modeled application during its run,
contention from resource sharing, etc.) \textit{at a given time}. $\omega$ represents randomness acting on the
system.
The system $\zeta$ can be further decomposed as:
\begin{equation}
    \zeta = \zeta_g(t) + \zeta_l(t, j)
\end{equation}
The component $\zeta_g(t)$ represents the \textit{global system impact} on all jobs running on the system (e.g., a
service degradation that equally impacts all jobs) and is only a function of time $t$. The component $\zeta_l(t, j)$
represents the \textit{local system impact} on the I/O throughput of job $j$ caused by resource contention and interactions with other jobs
running on the system. Contrary to the $\zeta_g(t)$ component, $\zeta_l(t, j)$ is job-specific and depends on the behavior of the
current set of applications running on the system, the sensitivity of $j$ to resource contention and noise, etc. Without
loss of generality, the I/O throughput of a job from Equation~\ref{eq:global_model} can be represented as:
\begin{equation} \label{eq:phi_breakdown}
\begin{split}
     \phi(j) &= f(j,\; \zeta_g(t),\; \zeta_l(t, j),\; \omega) \\
             &= f_a(j) + f_g(j, \zeta_g(t)) + f_l(j, \zeta_l(t, j)) + f_n(j, \zeta, \omega)
\end{split}
\raisetag{28pt}
\end{equation}

Here, $f_a(j)$ represents job throughput on an idealized system where the job is alone on the system, the system does not change over time, and there is no resource contention. $f_g(j, \zeta_g(t))$ represents how the evolving
configuration of the system (hardware provisioning, software updates, etc.) affects a job's I/O throughput. The $f_l(j,
\zeta_l(t, j))$ component represents the impact of resource contention and the sensitivity of $j$ to I/O noise. 
Finally, $f_n(j, \zeta, \omega)$ represents the impact of inherent system noise (e.g., packets dropped) on the job. 

\vspace{-0.17cm}
\subsection{Modeling assumptions}
The task of modeling a system's I/O throughput is to predict the behavior of the system when executing a job from some
application on some data. Modeling I/O throughput requires modeling both the HPC system and the jobs running on it.
Machine learning models used in this work attempt to learn the true function $\phi$ by mapping observable features of the
job $j$ and the system $\zeta$ to measured I/O throughputs $\phi(j)$. A model $m(j_o, \zeta_o)$ is tasked with predicting throughput $\phi(j)$, where $j_o$ and $\zeta_o$ are the observable job and system features.

When designing ML models, the choice of model architecture and model inputs has implicit assumptions about the process
that generates the data. When incorrect assumptions are made about the domain, the model will suffer errors that cannot
be fixed within that modeling framework. We investigate four common assumptions about the HPC domain.

\textbf{All data is in-distribution:}
a common assumption that practitioners make is that all model errors are the product of insufficiently trained models,
inadequate model architectures, or missing discriminative \textit{features}. However, some jobs in the
dataset may be \textit{Out of Distribution (OoD)}, that is, they may be collected at a different time or environment, or
through a different process. The model may underperform on OoD jobs due to the lack of similar jobs in the training
set and not due to lack of insight (features) into the job. The cause of the problem is \textit{epistemic uncertainty (EU)}
- the model suffers from lack of knowledge or \textit{reducible uncertainty}, since a broader
training set would make the OoD jobs in-distribution (ID). In the HPC domain, epistemic uncertainty is present in
cases of rarely run or novel jobs or uncommon system states.  Without considering the possibility that a portion of the
error is a product of epistemic uncertainty, practitioners will put effort into tuning models instead of collecting more
underrepresented jobs. Referring to Equation~\ref{eq:global_model}, this assumption may be expressed as: deployment
time $j_d$ and $\zeta_d$ are drawn from a different distribution from training time $j_t$ and $\zeta_t$.

\textbf{Noise is absent:} 
all systems have some inherent noise that cannot be modeled and will impact predictions.
\textit{Aleatory uncertainty} (AU) refers to inherent uncertainty, either due to noise or missing features about jobs in the
dataset.
Modeling errors due to aleatory uncertainty are
different from epistemic uncertainty because collecting more jobs may not reduce AU. HPC I/O domain experts note that
certain systems do have significantly higher or lower I/O noise~\cite{wan1, xie2}, but quantifying noise in modern leadership computing systems is an under-explored subject. Understanding and characterizing system's inherent I/O noise is key to account for the aleatory uncertainty
in the ML-based I/O models and better quantify the effect of this uncertainty on the I/O throughput predictions.  I/O
modeling works rarely attempt to quantify ML model uncertainty~\cite{madireddy_vae} even though an estimate of AU
significantly helps in model selection. The assumption that noise is not present in the dataset can be expressed as
follows: The practitioner assumes that the process has the form of $\phi(j) = f(j, \zeta)$ instead of $\phi(j) = f(j,
\zeta, \omega)$. 


\textbf{Sampling is independent:}
running a job on a system can be viewed as sampling the combination of system state and application
behavior and measuring I/O throughput.
Most I/O modeling works implicitly assume that multiple samples taken at the same time are independent of each other. The system is modeled as equally affecting
all jobs running on it, that is, the placement of different jobs on nodes, the interactions between neighboring jobs,
network contention, etc. \textit{do not affect the job}. This assumption can then be expressed as: the process has the
form of $\phi(j) = f(j, \zeta_g(t), \omega)$, not $\phi(j) = f(j, \zeta, \omega)$. 

\textbf{Process is stationary:}
a common assumption is that the data-generating process is stationary, and that the same job ran at
different times achieves the same I/O throughput. As hardware fails, as new nodes are provisioned, and
shared libraries get updates, the system evolves over time. This assumption is therefore incorrect, and ignoring it,
e.g., by not exposing \textit{when} a job is ran to the ML model may cause hard-to-diagnose errors. In other words, the
assumption is that the process has the form of $\phi(j) = f(j, \omega)$ and $f_g(j, \zeta_g(t)) = 0$.

\section{Classifying I/O throughput prediction errors}\label{sec:taxonomy}
No matter the problem to which machine learning is applied, a systematic characterization of
the sources of errors is crucial to improve model accuracy.
While there is no substitute for `looking at the data' to understand the root
cause of the problem, this approach does not scale for large datasets. We seek a systematic way to understand the barriers
to greater accuracy and improve ML models applied to system data. 

The key questions we ask in this work are: What are the impediments to the successful application of learning algorithms in
understanding I/O? Should ML practitioners focus on acquiring more data on HPC applications or the HPC system?
How much of the error stems from poor ML model architectures? How much of the error can be attributed to the dynamic
nature of the system and the interactions between concurrent jobs?  What fraction of jobs exhibit a truly novel I/O
behavior compared to the jobs observed thus far? At what point are the applications \textit{too novel}, so much so that
users should no longer trust the predictions of the I/O model?  We now describe five classes of errors and
dive deeper into error attribution in Sections~\ref{sec:stationary_errors}, \ref{sec:nonstationary_errors},
\ref{sec:data_collection_errors} and \ref{sec:generalization_errors}.

\begin{figure*}[t]
    \includegraphics[width=1.0\textwidth]{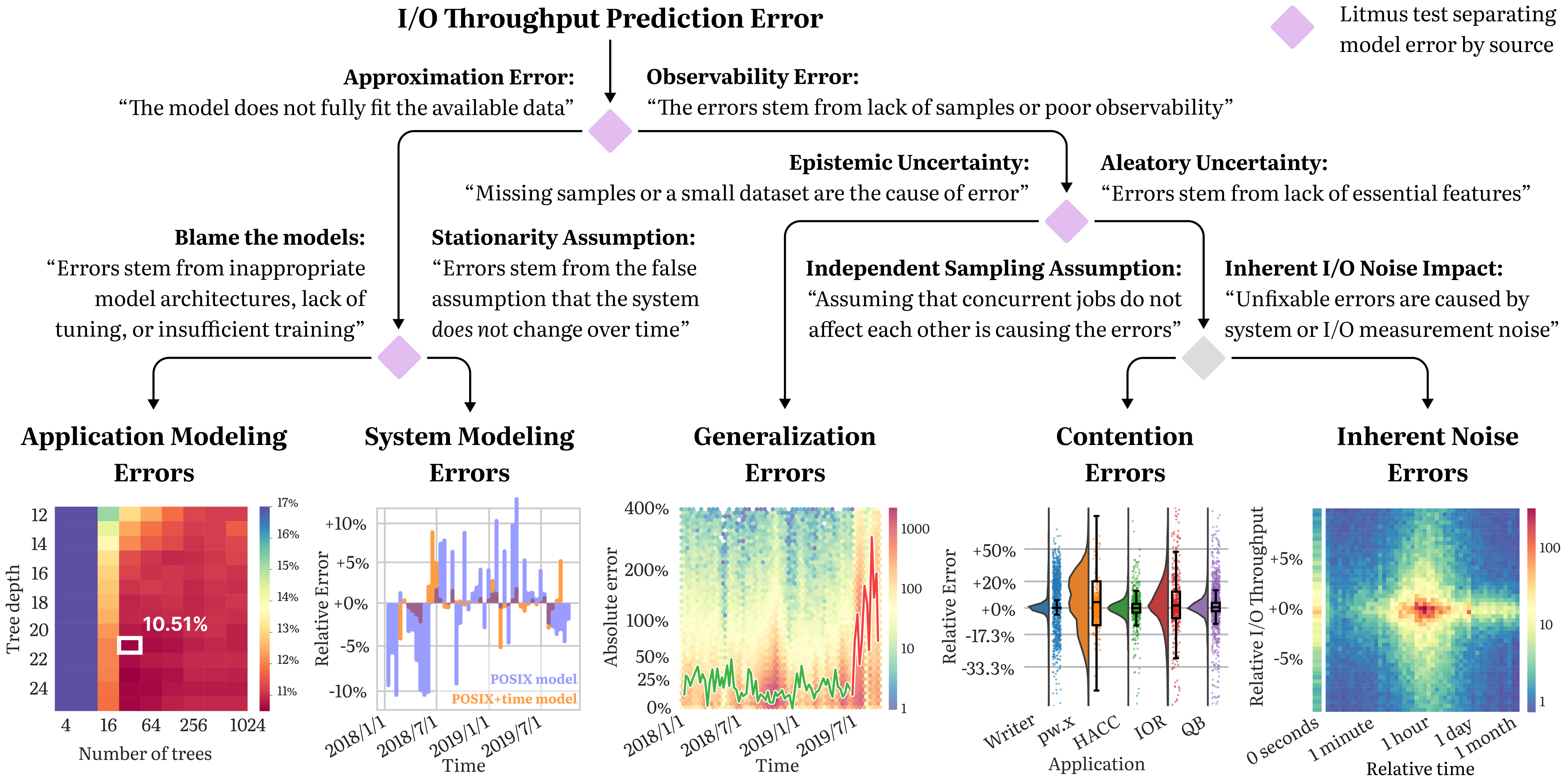}
    \caption{Taxonomy of I/O throughput modeling errors, with examples of the effects of each class of error - (a) Median errors of XGBoost models with varying numbers of estimators and estimator depth.(b) I/O throughput prediction error for sets of identical (duplicate) jobs, for 5 different applications. (c) Distribution of I/O throughput prediction errors and time ranges for sets of duplicate jobs. (d) Median errors before (green) and after training (red), with error distribution shown in the back.}
    \label{fig:teaser}
    \vspace{-0.2in}
\end{figure*}

The lack of application and system observability, the inherent noise $\omega$, and 
the OoD jobs prevent ML models from fully capturing system behavior, causeing errors. We define the I/O
throughput prediction error of a model $m$ in a job $j$ as: 
\begin{equation} 
    e(j) = \phi(j, \zeta, \omega) - m(j_o, \zeta_o)
\end{equation}
Following the $\phi(j)$ terms from Eq.~\ref{eq:phi_breakdown} and including the out-of-distribution
error, the error can be broken down as follows: 
\begin{equation}\label{eq:error_breakdown}
    e(j) = e_{app} + e_{system} + e_{OoD} + e_{contention} + e_{noise}
\end{equation}
Here, the application modeling error $e_{app}$ is caused by a poor model fit of application behavior ($f_a(j)$ component),
the global system error $e_{system}$ is caused by poor predictions of system dynamics ($f_g(j, \zeta_g(t))$ component),
the out-of-distribution error $e_{OoD}$ is caused by weak model generalization on novel applications or system states,
the contention error $e_{contention}$ is caused by poor predictions of job interactions ($f_l(j, \zeta_l(t, j))$ component),
and the noise error $e_{noise}$ is caused by the inability of any model to predict inherent noise ($f_4(j, \zeta, \omega)$ component). 
This leaves us with five classes of errors shown at the bottom of Figure~\ref{fig:teaser}. While attributing such an error
on a per-job basis is difficult, we will show that estimating each component across a whole dataset is possible.

\vspace{-0.15cm}
\subsection{I/O Model Error Taxonomy and Litmus Tests}
Errors in Equation~\ref{eq:error_breakdown} must be estimated in a specific order shown in Figure~\ref{fig:teaser} due
to the specifics of individual litmus tests. For example, before the effect of aleatory and epistemic
uncertainty can be separated, a good model must be found~\cite{autodeuq}. Similarly, before global and local system
modeling errors can be separated, OoD jobs must be identified. 


\textbf{Application modeling errors: }
ML models can have varying expressivity and may not always have the correct structure or enough parameters to fit the
available data. Models whose structure or training prevents them from learning the shape of the data-generating process
are said to suffer from \textit{approximation errors}, which are further divided into \textit{application} and \textit{system
modeling errors}. 
Application modeling errors are caused by poor predictions of application behavior which can be fixed through better I/O models or hyperparameter searches. 
The first column of Figure~\ref{fig:teaser} illustrates the impact of application modeling errors
with an example hyperparameter search over two XGBoost parameters on the Theta dataset. 
The search finds that 32 trees with depth of 21 perform best, while the XGBoost defaults use 100 trees of depth 6.
Approximation errors cannot be classified
as epistemic or aleatory in nature, because no new features or jobs are necessary to remove this error. To
estimate AU and EU in the dataset, methods such as AutoDEUQ~\cite{autodeuq} require first that an appropriate model
architecture is found and trained, so approximation errors are the first branch of the taxonomy. 

\textbf{System modeling errors:}
system behavior changes over time due to transient or long-term changes such as filesystem metadata
issues, failing components, new provisions, etc.~\cite{lockwood_pdsw17}. A model that is only aware of application
behavior, but not of system state implicitly assumes that the process is stationary. It will be forced to learn the
\textit{average} system response to I/O patterns, and will suffer greater prediction errors during periods when system
behavior is perturbed. Errors that occur due to poor modeling of the global system component $\zeta_g(t)$ are called
\textit{system modeling errors}. To illustrate this class of errors, two models are trained to predict I/O throughput in the second column of Figure~\ref{fig:teaser}, and each model's weekly average error is plotted against time.
The blue model is exposed only to application behavior, while the orange model also knows the \textit{job start time}.
During service degradations, the blue model has long periods of biased errors while the orange model does not, since it knows when degradations happen.

\textbf{Generalization errors: }
ML models generally perform well on data drawn from the same distribution from which their training set
was collected. When exposed to samples highly dissimilar from their training set, the same models tend to make
mispredictions. These samples are called `out-of-distribution' (OoD) because they come from new, shifted,
distributions, or the training set does not have full coverage of the sample space.
As an example, the third column of Figure~\ref{fig:teaser} shows model error before (green) and after (red) deployment, with the error
significantly rising when the model is evaluated on data collected outside the training time span.

\textbf{Contention errors:}
a diverse and variable number of applications compete for compute, networking, and I/O
bandwidth on HPC systems and interact with each other through these shared
resources~\cite{10.1145/3322789.3328743, 7877142}. Although the global system state will impact all jobs equally, the impact of resource sharing is specific to pairs
of jobs that are interacting and is harder to observe and model. Prediction errors that occur due to lack of
visibility into job interactions are called \textit{contention errors} and are shown in the fourth column of
Figure~\ref{fig:teaser}. Here, the I/O throughputs of a number of identical runs (the same code and data) of different
applications illustrate that some applications are more sensitive to contention than others, even when accounting for
global system state.

\textbf{Inherent noise errors:}
while hard to measure, resource sharing errors can potentially be removed through greater insight into the system and
workloads. What fundamentally cannot be removed are \textit{inherent noise errors}: errors due to random behavior by the
system (e.g., dropped packets, randomness introduced through scheduling, etc.). Inherent noise is problematic both
because ML models are bound to make errors on samples affected by noise and because noisy samples may impede model
training. The fifth column of Figure~\ref{fig:teaser} shows the I/O throughput and start time differences between pairs
of identical jobs. The leftmost column contains jobs that ran at exactly the same time, which often experience 5\%
or more difference in I/O throughput.
\vspace{-0.2cm}
\section{Datasets and experimental setup}\label{sec:preliminaries}
This work is evaluated on two datasets, one collected from ALCF Theta
supercomputer in the period from 2017 to 2020, and one collected from NERSC Cori
supercomputer in the period from 2018 to 2019. Theta collects Darshan~\cite{darshan} and Cobalt logs and consists
of about 100K jobs with an I/O volume larger than 1GiB, while Cori collects Darshan and Lustre Monitoring
Tools (LMT) logs, and consists of 1.1M jobs larger than 1GiB. 

Darshan is an HPC I/O characterization tool that collects HPC job I/O access patterns on both the POSIX and MPI-IO levels, and serves as our main insight
into application behavior. It collects POSIX aggregate job-level data, e.g., total number of bytes transferred, accesses made,
read / write ratios, unique or shared files opened, distribution of accesses per access size, etc. 
MPI-IO is a library built on top of POSIX that offers
higher-level primitives for performing I/O operations and may offer the model more insight into application behavior.
Darshan collects MPI-IO information for jobs that use it, and all requests through MPI-IO are also visible on the POSIX level.
The Cobalt scheduler logs number of nodes and cores assigned to a job, job start and end times, job placement, etc.
Cobalt logs may be useful to the model since the number of cores a job allotted is not visible to Darshan. Darshan
only collects the number of job processes, which is commonly equal to or greater than the number of cores allocated to a job.
LMT collects I/O
subsystem information such as storage server load and file system utilization, and serves as our main insight into the
I/O subsystem state over time. 
LMT records the state of object storage servers (OSS) and
targets (OST), and metadata servers (MDS) and targets (MDT) of the Lustre scratch filesystem every 5 seconds. Some of
the features LMT collects are CPU and memory utilization of the OSS's and MDS's, number of bytes
transferred to and from the OSTs, the fullness of the system, or the number of metadata operations (e.g., \texttt{open},
\texttt{close}, \texttt{mkdir}, etc.) performed by the metadata targets, etc. 
LMT collects per-OSS/OST/MDS/MDT logs, but since a job may be served by an arbitrary number of these I/O nodes, only the
minimum, maximum, mean and standard deviation are exposed to the ML model.
Overall, models have access to 48 Darshan POSIX, 48 Darshan MPI-IO, 37 LMT, and 5 Cobalt features. 


The ML models in this work are trained using supervised learning on the task of predicting the I/O throughput of
individual HPC jobs. The error models are optimizing is:
\begin{equation}
    e(y, \hat y) = \frac{1}{n} \sum^n_{i=1}\left|log_{10} \left(\frac{y_i}{\hat y_i}\right)\right|
\end{equation}
where $y_i$ and $\hat y_i$ are the $i$-th job's measured and predicted I/O throughputs. 
Because $log(x) = -log(1/x)$, if a model overestimates or underestimates the I/O throughput by the same
relative amount, the error remains the same. We use percentages to write errors, where, e.g., a -25\% error specifies 
that the model underestimated real I/O throughput by 25\%; however, some figures show the absolute error when model bias
is not important. While models try to lower mean error, they report median values since some of the distributions have heavy tails that make mean estimates unreliable.

\section{Application modeling errors}\label{sec:stationary_errors}
When an ML practitioner is tasked with a regression problem (predicting continuous values), the first model they
evaluate will likely under-perform on the task, e.g., due to inadequate data preprocessing, architecture, or
hyperparameters. Therefore, the model will suffer from \textit{approximation errors}, which can be removed by tuning the
model hyperparameters. Because inappropriate models may be unable to fit even easy tasks, this class of errors should
be solved before seeking, e.g., additional samples or sample features.

Figure~\ref{fig:teaser} shows that the approximation consists of the application and
system approximation. This section asks whether I/O models build faithful representations of application
behavior. 
 The questions we
ask are as follows: What are the limits of I/O application modeling? In practice, do I/O models faithfully learn application
behavior? Can I/O application modeling benefit from extra hyperparameter fine-tuning or new application features? 

\subsection{Estimating limits of application modeling}
We develop an application modeling error litmus test whose objective is to separate the application modeling error $e_{app}$ from
the four other error classes in Equation~\ref{eq:error_breakdown}. To do so, we seek a `golden model' that models
application behavior as accurately as possible given the inputs. When practical ML models are compared with this golden model, an
estimate of the application modeling error can be made. 

To build this `golden model', we rely on a property of synthetic datasets where the data-generating process can be freely
and repeatedly sampled. When analyzing HPC logs, it is common to see records of the same application ran multiple times
on the same data, or data of the same format. For example, system benchmarks such as IOR~\cite{ior} may be run
periodically to evaluate filesystem health and overall performance. We call these sets of repeated jobs
\textit{`duplicate jobs'}. Jobs are duplicates if they belong to the same application and all their
\textit{observable} application features are identical.
Note that an application can have many different sets of duplicate jobs, likely ran with different input parameters.
Because the duplicate features fed to an ML model are identical, the model cannot distinguish between duplicates from
the same set, and will achieve best possible accuracy on the training set if it learns to predict the mean I/O
throughput value of a set.
A model that does not learn to predict a set's mean value suffers from application-modeling error.

Sets of duplicate jobs can be used to build a litmus test that evaluates the median absolute error of a model which has $e_{app}=0$. Any practical model with a median error $e^p$ 
can then learn its application modeling error $e^p_{app}$ by comparing it with the median error
of the `golden model' $e^g$ as $e^p_{app} = e^p - e^g$.  The litmus test is administered as: 
\begin{tcolorbox}[left=0mm,right=0mm]
\vspace{-0.1cm}
1. Sets of duplicate jobs in the dataset are found; 2. The mean I/O throughput of each set is calculated; 
3. This mean is subtracted from each duplicate's I/O throughput in the set to calculate duplicate error, and Bessel's correction is applied~\cite{bishop};
4. The median error across all duplicates is reported. 
\vspace{-0.1in}
\end{tcolorbox}
This average difference represents
the smallest average error 
that a model $m(j)$ can achieve on \textit{duplicate jobs}. Assuming that duplicate jobs are drawn
from the same distribution of applications as the whole dataset, the duplicate median absolute error represents the lower bound on
median absolute error a model can achieve on the whole dataset. Note that different applications may have different distributions of
duplicate I/O throughputs, as shown in the fourth column of Figure~\ref{fig:teaser}. 
For this litmus test to be accurate, a large sample of applications representative of the HPC system workload 
must be acquired. When applied to Theta, 19010 duplicates (23.5\% of the dataset) over 3509 sets show a median 
absolute error of 10.01\%. Cori has 504920 duplicates (54\%) in 77390 sets show a median absolute error of 14.15\%.
If the litmus test is correct, a model that has the same (but not lower!) median absolute error on the general dataset can be found.


\subsection{Minimizing application modeling error}
The next question is whether ML models can practically reach the estimate of the lower bound of the error. Several I/O modeling works
have explored different types of ML models: linear regression~\cite{isakov_sc20}, decision
trees~\cite{10.1007/978-3-319-58667-0_19}, gradient boosting machines~\cite{isakov_sc20, 10.1007/978-3-319-58667-0_19, xie1},
Gaussian processes~\cite{10.1007/978-3-319-92040-5_10}, neural networks~\cite{10.1145/3337821.3337922}, etc. Here, we explore two types of models: XGBoost~\cite{xgboost}, an implementation
of gradient boosting machines, and feedforward neural networks. These model types are chosen for their accuracy,
scalability, and previous success in I/O modeling. 

Neither type of model achieves ideal performance `out of the box'. XGBoost model performance can be improved through
hyperparameter tuning, e.g., by exploring different (1) numbers of decision trees, (2) their depth, (3) the features each
tree is exposed to, and (4) part of the dataset each tree is exposed to. Neural networks are more complex, since they
require tuning hyperparameters (learning rate, weight decay, dropout, etc.), while also exploring different
architectures (number, size, type of layers, and their connectivity). In the case of XGBoost, we exhaustively explore four
hyperparameters listed above, for a total of 8046 XGBoost models. 
In the case of neural
networks, exhaustive exploration is not feasible due to state space explosion, so we use
AgEBO~\cite{DBLP:journals/corr/abs-2010-16358}, a Network Architecture Search (NAS) method that trains populations of
neural networks and updates each subsequent generation's hyperparameters and architectures through Bayesian logic.

The leftmost column of Figure~\ref{fig:teaser} shows a heatmap of an exhaustive search over two parameters on the Theta 
dataset, with the other two parameters (\% of columns and rows revealed to the trees) selected from the
best possible result found.  
The best performing model has an error rate of 10.51\% - very close to the predicted bound.
The search on Cori data arrives at a similar configuration (omitted due to space).

\begin{figure}[h]
    \vspace{-0.3cm}
    \centering
        \includegraphics[width=\columnwidth]{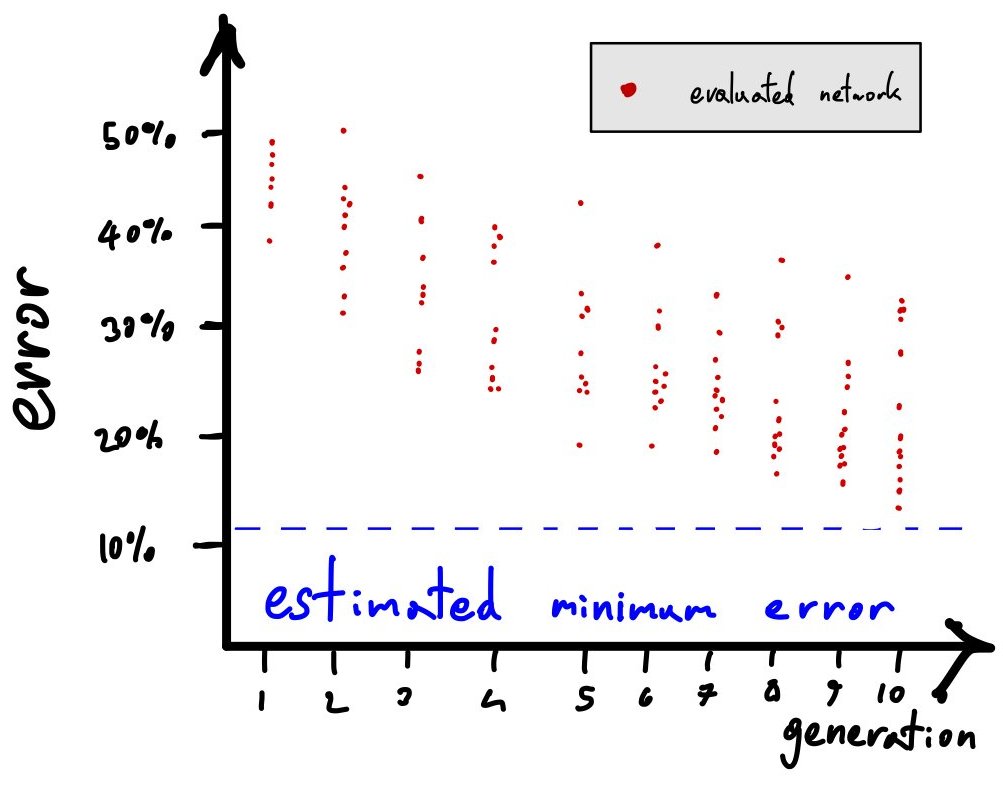}
    \caption{Results of the Neural Architecture Search (NAS), with the estimated error lower bound highlighted in red.}
    \label{fig:dnn_nas}
    \vspace{-0.3cm}
\end{figure}

In the case of neural networks, Figure~\ref{fig:dnn_nas} shows a scatter plot of test set errors of 10 generations
of neural networks on the Cori system, with 30 networks per generation. The networks are evolved using a separate
validation test to prevent leakage of the test set into the model parameters. Networks approach the estimated error
limit, and the best result achieves a median absolute error of 14.3\%. The fact that (1) after extensive tuning both
types of models asymptotically approach the estimated limit in model accuracy, and (2) NAS does little to improve
models, since models improve predictions only 6 times (gold stars). This suggests that both types of ML models
are impeded by the same barrier and that the architecture and the tuning of models are not the fundamental
issue in achieving better accuracy, i.e., that the source of error lies elsewhere.

\subsection{Increasing visibility into applications}
While hyperparameter and architecture searches approach but do not surpass the litmus test's estimated lower bound on
error, this is not conclusive evidence that all application modeling error has been removed and that error stems
from other sources. Possibly, there exist missing application features that might further reduce errors. We explore
two such sets of features: MPI-IO logs and Cobalt scheduler logs. 

\begin{figure}[h]
    \centering
    \includegraphics[width=0.85\columnwidth]{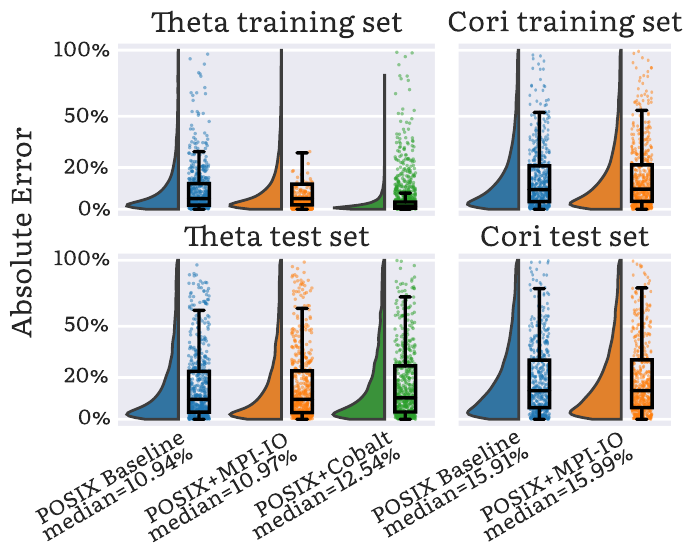}
    \vspace{-0.0in}
    \caption{Error distributions for models trained on POSIX, POSIX + MPI-IO, and POSIX + Cobalt feature
    sets.} 
    \label{fig:stationary_models}
    \vspace{-0.2in}
\end{figure}

Figure~\ref{fig:stationary_models} shows the absolute error distribution of the tuned models on three Theta datasets:
POSIX, POSIX + MPI-IO, and POSIX + Cobalt (Cori excluded because of the lack of Cobalt logs). None of the dataset enrichments
help reduce error, corroborating the conclusion that application modeling is not a source of error for these models, and further insight into applications will not help. Adding Cobalt logs does reduce the error on
the training set, and our experiments show that the job start and end time features are the cause.
Once timing features are present in the dataset, no two jobs are duplicates due to small timing variations.  Although
previously the ML model was not able to overfit the dataset due to the existence of duplicates, this is no longer the
case, and the ML model can differentiate and memorize each individual sample. In~\cite{isakov_sc20} authors remove
timing features for a similar reason: ML models can learn Darshan's implementation of I/O throughput calculation and
achieve good predictions without having to observe job behavior.

\section{Global system modeling errors}\label{sec:nonstationary_errors}
The second part of the approximation error in our taxonomy is the global system modeling error. This error 
refers to I/O climate and I/O weather effects~\cite{lockwood_pdsw17} that affect all jobs running on the system, and
corresponds to the second component in Equation~\ref{eq:phi_breakdown}. While global and local system impacts have
complex and overlapping effects, we suggest that factorizing the impact applied to all jobs versus the impact that is
dependent on pairs of concurrent jobs is useful for modeling purposes. 
The main difference between the two is that local system impacts cannot be predicted or modeled without knowledge of all
jobs running on the system, while global system impacts can. In other words, global system impacts can be expressed as a
property of the system at a given time, effectively compressing a part of system behavior.
We now ask: How does I/O contention impact job I/O throughput prediction? What are the
limits of global system modeling? How can I/O models approach this limit? 

\subsection{Estimating limits of global system modeling}
Global system impact $\zeta_g(t)$ on job $j$ from Equation~\ref{eq:phi_breakdown} can be formalized as some
function $\zeta_g(t) = g(J(t))$ where $J$ is the set of jobs running at time $t$. Since jobs have a start and end
time, given a dataset with a dense enough sampling of $J$, $g(J(t))$ can be calculated for every point in time. During periods of time where
the file system is suffering a service degradation, all jobs on the system will be impacted with varying severity.  A
model of the system does not need to understand how and why the degradation happened, it only needs to know the period
when it lasted and how different types of jobs were impacted. This time-based model is useless for predicting
future performance, and it's only utility is in evaluating how much of the degradation can be described as purely a
function of time. A deployed model does not have data on the future and will still need to observe the system.

To evaluate the global system impact, a golden model that exhibits no global modeling error is developed, against which we
can compare other, `real' ML models. Since the global system impact of $\zeta_g(t)$ only depends on time and does not need the set of all jobs $J$, only application behavior $j$ and the job start time
feature are exposed to the golden model. Here, a golden model is an XGBoost model fine tuned on a validation set and
evaluated on a test set. Assuming that the golden model is exposed to enough jobs throughout the lifetime of the system,
it will learn the impact of $\zeta_g(t)$ even without having access to the underlying system features causing that
impact. This golden model is used in the following litmus test:
\begin{tcolorbox}[left=0mm,right=0mm]
\vspace{-0.1cm}
1. The timing feature is
added to the Darshan-only (no Lustre or Cobalt) dataset; 2. A hyperparameter search is performed on a validation set and
application modeling error is removed; 3. the final model is evaluated on the test set, and it's error is reported.
\vspace{-0.1in}
\end{tcolorbox}
If the litmus test is correct, the golden model only suffers from the last three classes of errors: poor generalization, 
local system impact, and inherent noise. 
In Figure~\ref{fig:lustre_model_comparison} we evaluate a baseline model (blue) and a model enriched with the
job start time (orange). The addition of this single feature has a large impact on error: on
Cori, the error drops $40\%$, from 16.49\% down to 10.02\%, while on Theta the error drops by 30.8\%. Note that to obtain high accuracy on the POSIX+start time, a much larger model is needed, i.e., one that can remember the I/O weather
throughout the lifetime of the system.  

\subsection{Improving modeling through I/O visibility}
\begin{figure}
    \centering
    \includegraphics[width=0.95\columnwidth]{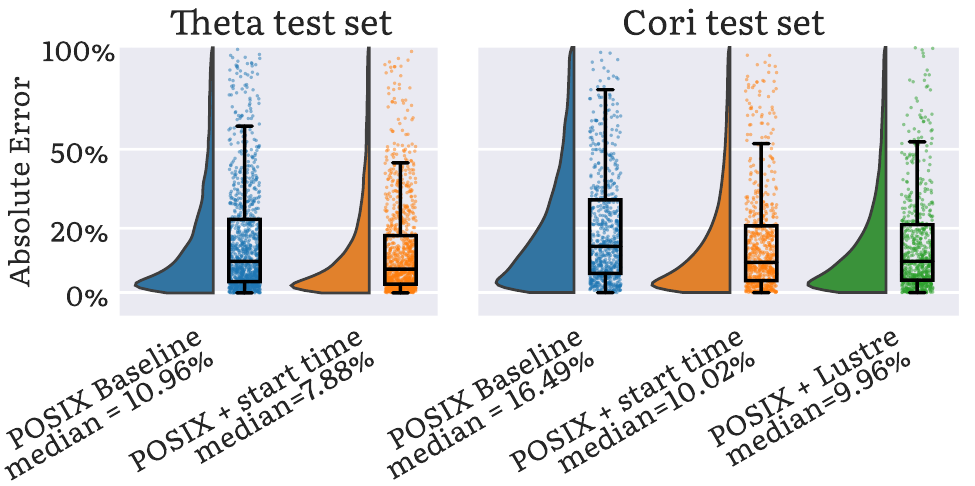}
    \caption{
    Error distribution of models trained on (1) POSIX, (2) POSIX + 
    the start time feature, and (3) Darshan and Lustre}
    \label{fig:lustre_model_comparison}
    \vspace{-0.5cm}
\end{figure}

With an estimate of minimal error achievable assuming perfect application and global system modeling, we investigate
whether I/O subsystem logs can help models approach this limit. Since Theta does not collect I/O subsystem logs, we
analyze Cori, which collects both application and I/O logs.  Figure~\ref{fig:lustre_model_comparison} shows the XGBoost
performance of three models: a baseline where $e_{app}=0$ (blue), the litmus test golden model where also $e_{system}=0$ (orange), and a
Lustre-enriched model (green). Cori's median  absolute error is reduced by 40\%, from
16.49\% down to 9.96\%. The Lustre-enriched results are surprisingly close to the litmus test's predictions,
and suggest that predictions cannot be improved through further I/O insight since the litmus test's prediction is
reached.

\section{Generalization errors}\label{sec:generalization_errors}
The remaining three classes of error are caused by lack of data and not poor modeling, as the top branch of the
taxonomy shows. While I/O contention and inherent noise errors are examples of aleatory uncertainty and
are caused by lack of insight into specific jobs, generalization errors stem from epistemic uncertainty, i.e., the lack
of other logged jobs around a specific job of interest. 
To motivate this section, in the third graph of
Figure~\ref{fig:teaser} we show error distribution of a model trained on data from January 2018 to July 2019. When 
evaluated on held-out data from the same period, the median absolute error is low (green line). Once the model is
deployed and evaluated on the data collected after the training period (July 2019 and after), median error spikes up (red
line).

\subsection{Estimating generalization error}

\begin{figure}[h]
    \vspace{-0.2cm}
    \centering  
    \includegraphics[width=0.95\columnwidth]{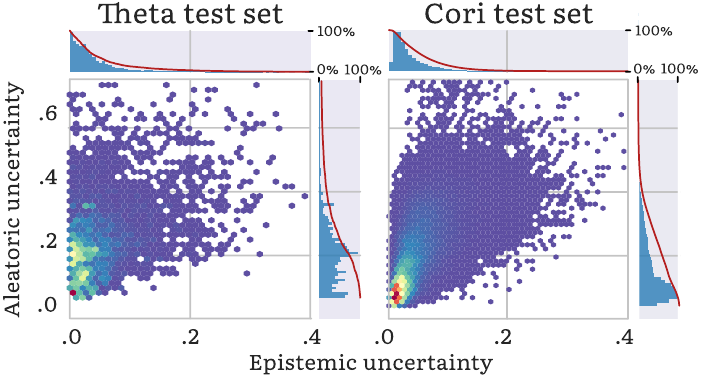}
    \caption{Distribution of job aleatory and epistemic uncertainties for the two systems, with marginal distributions
    and inverse cumulative error shown on the margins.} 
    \label{fig:autodeuq}
    \vspace{-0.2cm}
\end{figure}

Estimating the amount of out-of-distribution error $e_{OoD}$ is important because any
unaccounted OoD error will be classified as noise or contention. This will make systems that run a lot of novel jobs
appear to be more noisy than they truly are. Because OoD and ID jobs will likely have a similar amount of I/O and
contention noise, it is better to have false positives (ID jobs classified as OoD) than the other way, since false
negatives contribute to overestimating I/O noise.
To estimate the impact of out-of-distribution jobs on error $e_{OoD}$, we aim to quantify the how much of error is
epistemic, and how much is aleatory in nature, as shown in Figure~\ref{fig:teaser} (upper right).
The leading paradigm for uncertainty quantification works by training an ensemble of models and evaluating all of the
models on the test set. If the models make the same error, the sample has high aleatory uncertainty, but if the models
disagree, the sample has high epistemic uncertainty~\cite{10.5555/3295222.3295387}. The intuition is that predictions on
out-of-distribution samples will vary significantly on the basis of the model architecture, whereas predictions on noisy
samples will exhibit the same amount of error. Since this method relies on ensemble model diversity, several works have explored
increasing diversity through different model hyperparameters~\cite{Wenzel2020HyperparameterEF}, different
architectures~\cite{Zaidi2020NeuralES}, or both~\cite{autodeuq}. We choose to use AutoDEUQ~\cite{autodeuq}, a method
that evolves an ensemble of neural network models and jointly optimizes both the architecture and hyperparameters of the
models. AutoDEUQ's Neural Architecture Search (NAS) is compatible with the NAS search from
section~\ref{sec:stationary_errors}, reducing the computational load of applying the
taxonomy. Figure~\ref{fig:autodeuq} shows the distribution of epistemic (EU) and aleatory uncertainties (AU) of Theta
and Cori test sets. For both systems, aleatoric uncertainty is significantly higher than epistemic uncertainty.
Furthermore, \textit{all} jobs seem to have AU larger than some about 0.05, hinting at the inherent noise present in the
system. The inverse cumulative distributions on the top and right show the what percentage of total error is caused by
AU / EU \textit{below} that value. For example, for both systems 50\% of all error is caused by jobs with EU below 0.04,
while in case of AU, 50\% of error is below AU=0.25. The low total EU is expected since the test set was drawn from the
same distribution as the training set, and increases on the 2020 set (omitted due to space concerns). 

Epistemic uncertainty does not directly translate into the out-of-distribution error $e_{OoD}$ from
Equation~\ref{eq:error_breakdown}. When a sample is truly OoD, it may not be possible to separate aleatory and epistemic
uncertainty, since a good estimate of AU requires dense sampling around the job of interest. Therefore, we choose to
attribute all errors of a sample marked as out-of-distribution to $e_{OoD}$. This error attribution requires classifying
every test set sample as either in- or out-of-distribution, but since EU estimates are continuous values, an EU
threshold which will separate OoD and ID samples is required. Although this threshold is specific to the dataset and
may require tuning, the quick drop or `shoulder' in inverse cumulative error around EU=0.1 in Figure~\ref{fig:autodeuq}
makes the choice of an $e_{OoD}$ threshold robust. 
A litmus test that estimates the error due to out-of-order samples has the following steps:
\begin{tcolorbox}[left=0mm,right=0mm]
\vspace{-0.1cm}
1. Run NAS and collect the best performing models; 2. Estimate the aleatory and epistemic uncertainty using AutoDEUQ; 3. Find a stable EU threshold and classify the samples as ID and OoD; 4. Calculate $e_{OoD}$ as the sum of OoD sample errors. 
\vspace{-0.1in}
\end{tcolorbox}
On Theta, for an EU threshold of 0.24, .7\% of the samples are classified as OoD, but constitute 2.4\%
of the errors, while on Cori 2.1\% of error gets removed for the same EU threshold. In other words, the selected jobs have $3\times$ larger average error than random samples. By manually exploring the types of jobs that do get removed, we confirm that these are typically rare or novel applications.

\section{I/O Contention and Inherent Noise Errors}\label{sec:data_collection_errors}
With the ability to estimate the amount of application and system modeling error, as well as detect outlier jobs,
leftover error is caused by system contention or inherent noise. Both of these error classes are caused by aleatory
uncertainty, since the model lacks deeper insight into jobs or the system, as opposed the OoD case where the model lacks
samples. While e.g., application error was predictable and explainable in terms of broad application behavior (e.g.,
this application is slow because it frequently writes to shared files), the impact of contention and noise on I/O
throughput is caused by lower level, transient effects. Though it may be possible to observe and log such effects
through microarchitectural hardware counters or network switch logs, such logging would require vast amounts of storage
per job and would impact performance. Lack of practical logging ability makes the last two error categories typically
\textit{unobservable}. Furthermore, these two classes may only be separated in hindsight, and while I/O noise levels
may be constant, the amount of I/O contention on the system is unpredictable for a job that is about to run.

The question we ask in this section are: how can errors due to noise and contention be separated from errors due to poor
modeling or epistemic uncertainty? Is there a fundamental limit to how accurate I/O models can become? Can system I/O
variability be quantified? 

\subsection{Establishing the bounds of I/O modeling}
To separate the contention and noise from the first three classes of error, we develop a litmus test based on the test
from Section~\ref{sec:stationary_errors}. 
There, by observing sets of duplicates, the performance of an ideal model was estimated for which $e_{app} = 0$. Comparing real models against this ideal model allowed us to calculate real model's $e_{app}$.
A similar litmus test can be designed estimate the sum of contention and noise error, where only concurrent duplicates are
observed and both application behavior $j$ and global system behavior $\zeta_g$ can be held static for each duplicate set:
\begin{tcolorbox}[left=0mm,right=0mm]
\vspace{-0.1cm}
1. OoD jobs are removed and sets of duplicate jobs ran at the same time ($\Delta t=0$) are collected;
2. The mean I/O throughput of each set is calculated; 
3. Duplicate error is calculated as before;
4. The median error across all duplicates is reported. 
\vspace{-0.1in}
\end{tcolorbox}

\begin{figure}[h]
    \centering
    \vspace{-0.5cm}
    \includegraphics[width=0.8\columnwidth]{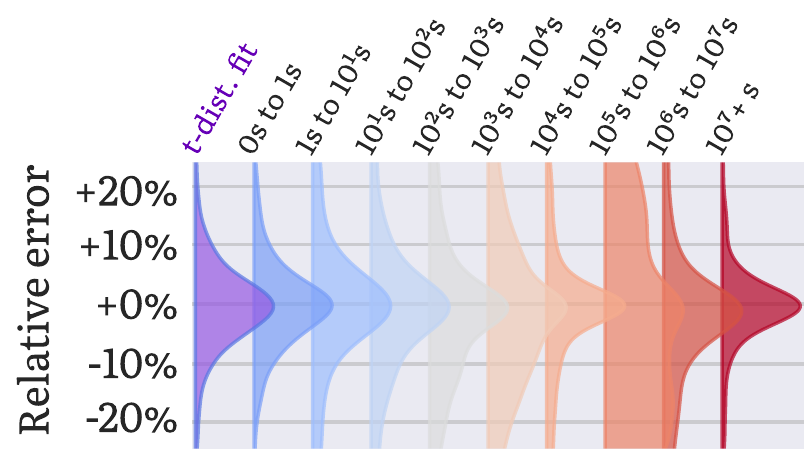}
    \caption{Distribution of errors for different periods between duplicate runs.}
    \label{fig:duplicates_over_time_kde}
    \vspace{-0.2cm}
\end{figure}

In the fifth column of Figure~\ref{fig:teaser} we show the distribution of I/O throughput differences $\Delta \phi$ and
timing differences $\Delta t$ between all pairs of Cori duplicate jobs, weighted so that large duplicate sets are not
overrepresented. The vertical strip on the left contains Cori duplicate jobs that were ran
simultaneously, largely because they were batched together. These jobs share $j$ and $\zeta_g$, but may differ in
$\zeta_l$ and $\omega$. Due to the denser sampling around 1 minute to 1 hour range, it is not immediately apparent
how the I/O difference changes between duplicates ran at the same time and duplicates ran with a delay. By grouping
duplicates from different $\Delta t$ ranges and independently scaling them, a better understanding of duplicate I/O
throughput distributions across timescales can be made, as shown in Figure~\ref{fig:duplicates_over_time_kde}. For both
systems (Cori omitted due to lack of space), the distributions on the right contain jobs ran over large periods of time where
global system impact $\zeta_g$ might have changed, explaining the asymmetric shape of some of them. The left-most
distributions are similar, since variance only stems from contention $\zeta_l$ and noise $\omega$. While some
distributions (e.g., the $10^5$ to $10^6$ second) show complex multimodal behavior, all of the
distributions seem to contain the initial zero second ($0s$ to $1s$) distribution. 

By fitting a normal distribution to the $\Delta t=0$ distribution (0s to 1s) in Figure~\ref{fig:duplicates_over_time_kde}, we can both
(1) learn the lower limit on total modeling error and (2) learn the system's I/O noise level, i.e., how much I/O
throughput variance should jobs running on the system expect. However, upon closer inspection, the $\Delta t=0$
distribution \textit{does not} follow a normal distribution. This is surprising, since if noise is normally
distributed, independent over time, and it's effects are cumulative, the total impact is a sum of normal distributions,
which should also be a normal distribution. The answer lies in how the concurrent ($\Delta t=0$) duplicates are sampled.
When observing duplicates, in general, duplicate sets have between 2 and hundreds of thousands of identical jobs in them.
However, in duplicate sets with identical start times on Theta, 70\% of the sets only have two identical jobs, and 96\% have
6 jobs or less, with similar results on Cori. The issue stems from how small (sub-30 samples) duplicate set
errors are calculated: when only a small number of jobs exist in the set, the mean I/O throughput of the set is biased
by the sampling, i.e., the estimated mean is closer to the samples than the real mean is. This causes the set I/O
throughput variance to decrease and therefore duplicate error estimate will be reduced as well. Student's
\textit{t}-distribution describes this effect: when the true mean of a distribution is known, error calculations follow
a normal distribution. When the true mean is not known, the biased mean estimate makes the error follow the
\textit{t}-distribution. As the sample count reaches 30 and above, the \textit{t}-distribution approaches the normal
distribution.


Finally, we seek to estimate the I/O noise variance of a given system. Naively taking the variance of the
\textit{t}-distribution will produce a biased sample variance $\sigma^2$, but it can be fixed by applying Bessel's
correction as $\frac{n}{n-1}\sigma^2$. 
In practical terms, a job
running on Theta can expect an I/O throughput within $\pm5.71\%$ of the predicted value \textit{68\% of the
time}, or within $\pm 10.56\%$ 95\% of the time. For Cori, these values are $\pm 7.21\%$ and $\pm 14.99\%$, respectively.
This is a fundamental barrier not just to I/O model improvement, but to predictable system usage in general. Although some insight into contention can be gained through low-level logging tools, noise cannot be overcome. I/O practitioners
can use this litmus test to evaluate the noise levels of their systems, and ML practitioners should reconsider how they
evaluate models, since some systems may be simply harder to model. 

%

\section{Applying the taxonomy} \label{sec:framework}

\begin{figure*}
    \centering
    \includegraphics[width=.99\linewidth]{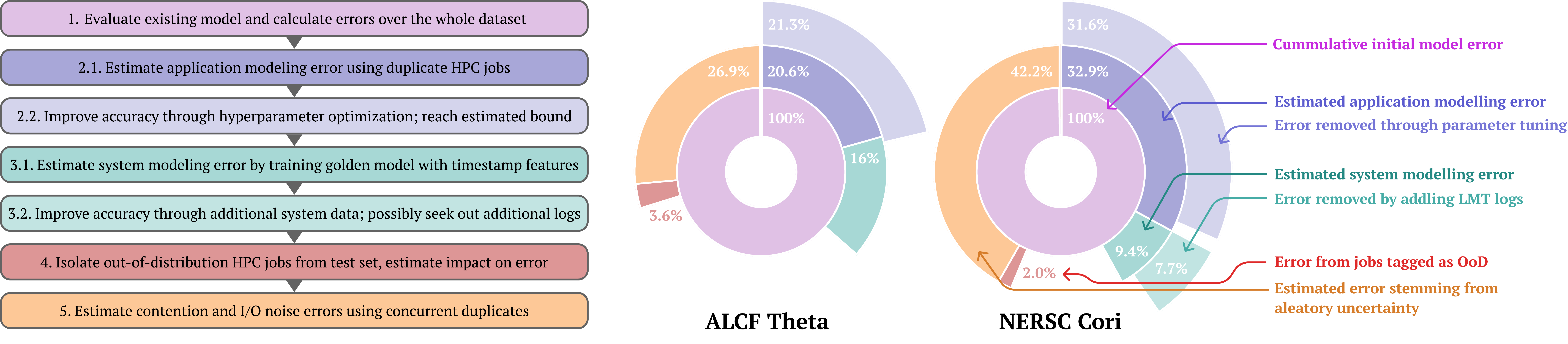}
    \caption{Framework for applying the taxonomy (left), and the results from ALCF Theta and NERSC Cori systems.} 
    \label{fig:framework}
    \vspace{-.2in}
\end{figure*}

We now illustrate how the proposed taxonomy can be used in practice. In Figure~\ref{fig:framework} (left), we show the
steps a modeler can follow to evaluate the taxonomy on a new system. 
Step 1: The modeler splits the available data into training and test sets, and then trains and evaluates some baseline
machine learning model on the task of predicting I/O throughput. This model does not have to be perfect, as the
taxonomy should reveal what are the main sources of error and approximately how much the quality of the model is at
fault. 
Step 2.1: The modeler estimates application modeling errors by finding duplicate jobs and evaluating the mean predictor
performance on every set of duplicates. Assuming that the distribution of duplicate HPC jobs is representative of the
whole population of jobs, this step provides the modeler with a lower bound on the application modeling error. 
Step 2.2: By contrasting the baseline model error (Step 1) and the estimated application modeling error, the modeler can
estimate the percentage of error that can be attributed to poor modeling. The modeler performs a hyperparameter or network
architecture search and arrives at a good model close to the bound.  
Step 3.1: The modeler estimates system modeling errors by exposing the job start time to a golden model.
This step requires that the modeler has developed a well-performing model in Step 2.2, that is, one that achieves
close to the estimated ideal performance. The test set error of the model serves as an estimate of the application
+ system modeling lower bound. 
Step 3.2: The modeler explores adding sources of system data to improve the performance of the baseline
model up to the estimated limit of application and system modeling. 
Step 4: The modeler identifies out-of-distribution samples using AutoDEUQ, calculates OoD error that stems
from these samples, and removes them from the dataset. 
%
Step 5: The modeler estimates the error that can be attributed to contention and noise, as well as I/O variance of the
system. This estimate is made by observing the I/O throughput differences between sets of concurrent duplicates, i.e.,
duplicate jobs ran at around the same time. 

In the middle and right portion of Figure~\ref{fig:framework} we show the average baseline model error (inner pink circle
segment) of both ANL Theta and NERSC Cori systems, and how that error is broken down into different classes of error.
We do not focus on the cumulative error value of the two systems; instead, we focus on attributing the baseline model error
into the five classes of errors in the taxonomy (middle circle segments of the pie chart), and how much improved
application and system modeling can help reduce the cumulative error (outer segments of the pie chart).  
The inner blue section of the two pie charts represents the estimated application modeling error,
as arrived at in Step 2.1. The outer blue section represents how much of the error can be fixed through hyperparameter
exploration, as explored in Step 2.2. 
The inner green section represents the estimated system modeling error, derived in Step 3.1. Note that the total
percentage of system modeling error is relatively small on both systems; i.e., I/O contention, filesystem health,
hardware faults, etc., do not have a dominant impact on I/O throughput. The outer green circle segment represents the
percentage of error that can be fixed by including system logs (LMT logs in our case), as described in Step 3.2. Only
the Cori pie chart has this segment, as Theta does not collect LMT logs. On Cori, the inclusion of LMT logs helps remove
most of the system modeling errors, pointing to the conclusion that including other logs (i.e., topology, networking)
may not help to significantly reduce errors. 
The inner red segment represents the percentage of error that can be attributed to out-of-distribution samples of the
two systems, as calculated in Step 4. 
Finally, the yellow circle segment represents the percentage of error that can be attributed to aleatory uncertainty.
For both Theta and Cori, this is a rather large amount, pointing to the fact that there exists a lot of innate noise in the
behavior of these systems, and setting a relatively high lower bound on ideal model performance. 

The similarity between the modeling error estimates (Steps 2.1 and 3.1) and the actual updated model performance (Steps 2.2
and 3.2) is surprising and serves as evidence for the quality of the error estimates. However, the estimates of the five error classes \textit{do not} add up to 100\%. The first three error estimates are just that - estimates,
derived from a subset of data (duplicate HPC jobs) that do not necessarily follow the same distribution as the rest of
the dataset and may be biased. If we add the estimates, we see that on Theta 32.9\% of the error is
unexplained, and on Cori 13.5\% of the error is unexplained.
Cori's lower unexplained error may be due to the fact that we have collected some 1.1M jobs compared to 100K on Theta. 

\vspace{-0.3cm}
\section{Discussion}~\label{sec:discussion}
Developing production-ready machine learning models that analyze HPC jobs and predict I/O throughput is difficult: the
space of all application behaviors is large, HPC jobs are competing for resources, and the system changes over time. To efficiently improve these models, we present a taxonomy of HPC I/O modeling errors that allows us to study the
types of errors independently, quantify their impact, and identify the most promising avenues for model improvement. Our
taxonomy breaks errors into five categories: (1) application and (2) system modeling errors, (3) poor generalization, (4) 
resource contention, and (5) I/O noise. We present litmus tests that quantify what percentage of existing error belongs
to the classes and show that models improved by using the taxonomy are within several percentage points of an
estimated best-case I/O throughput modeling accuracy. We show that a large portion of I/O throughput modeling error is irreducible and stems from I/O variability. We provide tests that quantify the I/O variability and establish an upper bound on how accurate models can become. Our test shows that jobs run on Theta and Cori can expect an I/O throughput standard deviation of 5.7\% and 7.2\%, respectively.

\bibliographystyle{IEEEtran}
\bibliography{paper}

\end{document}